\documentclass[10pt, twoside]{article}
\usepackage[b5j,hmargin={1 in,0.6 in},vmargin={1 in,0.6 in}]{geometry}
\usepackage{times}
\usepackage[labelsep=space, labelfont=bf]{caption}
\usepackage{amsmath,amssymb,amsthm,latexsym}
\makeatletter	
\g@addto@macro\th@definition{\thm@headpunct{}}
\g@addto@macro\th@plain{\thm@headpunct{}}
\g@addto@macro\th@remark{\thm@headpunct{}}
\makeatother
\usepackage{graphicx}
\usepackage{titlesec}
\titleformat{\section}{\normalsize\bfseries}{\hspace{0mm} \thesection.}{10 pt}{}
\usepackage{indentfirst}
\usepackage{setspace}
\usepackage[round]{natbib}
\usepackage{url}
\usepackage{hyperref}
\hypersetup{pdfborder={0 0 0}}   

\providecommand{\Abstract}[1]{\noindent\textbf{Abstract} \\ #1}
\providecommand{\titlepaper}[1]{\begin{spacing}{1.15}\noindent\bf\Large#1\end{spacing}}
\providecommand{\keywords}[1]{\noindent\rule{0.3 \textwidth}{1 pt }\newline \textbf{Keywords:} #1}
\theoremstyle{definition}
\numberwithin{equation}{section} 

\usepackage{mathrsfs} 	%%	Math. Character:	\mathscr{}
\usepackage{booktabs}

\DeclareMathOperator{\arcsec}{\arcsec\,}

\begin{document}
\thispagestyle{empty}
%% 	
%\logothailstat
%\vskip2cm
%%----------		title		----------%%
\titlepaper{%
Autoregressive Moving Average and Generalized Autoregresive Moving Average in Covid-19's Confirmed Cases in Indonesia%\\
%Manuscript title line 2 
}	%%

%%----------		name of the authors		----------%%
\noindent
%%
%%%%%		for only one author 
%%%
%Name1 Surname\textsuperscript{*}
%%%
%\vskip 1pt \noindent%%
%%%
%Affiliation Dept/Program/Center, Institution Name, City, State, Country
%%%
%%%%%
%%
%%%%%		for two authors
%%
\fontsize{11pt}{11pt}\textbf{Khusnia Nurul Khikmah, A'yunin Sofro\textsuperscript{*}}

\vskip 1pt \noindent\text{Mathematics Department, Universitas Negeri Surabaya, East Java, Indonesia}

%%
%%%%%
%%
%%%%%		for three authors
%%%
%Name1 Surname,\textsuperscript{\hspace{-3 pt}1,*} 
%Name2 Surname\textsuperscript{2} and
%Name3 Surname\textsuperscript{3}
%%%
%\vskip 1pt \noindent\textsuperscript{1}%%
%Affiliation Dept/Program/Center, Institution Name, City, State, Country
%%%
%\vskip 1pt \noindent\textsuperscript{2}%%
%Affiliation Dept/Program/Center, Institution Name, City, State, Country
%%%
%\vskip 1pt \noindent\textsuperscript{3}%%
%Affiliation Dept/Program/Center, Institution Name, City, State, Country
%%%
%%%%%
\vskip 1.15pt 
\noindent
\textsuperscript{*}Corresponding author; e-mail: ayuninsofro@unesa.ac.id

%%----------		abstract		----------%%
\Abstract{\hspace*{12 pt}	%%
Autoregressive moving average and generalized autoregressive moving average are often used in statistical modeling. This study using this method because the method uses data from the previous period to model the data for the current period. In addition, the technique is often used in data prediction. The familiar data used is count data. Count data is the data that most often cause data not to spread usually. Therefore, time series modeling, one of which is through arithmetic series, was developed. This study aims to obtain the best modeling results from positive confirmed cases of Covid-19 in Indonesia. They were getting the results from the best modeling for positive confirmed cases of Covid-19 in Indonesia based on the smallest Aikake's information criterion value.
}	%%

%%----------		keywords		----------%%
\vskip 5 pt

%%----------		contents		----------%%
%% 	section 1
\section{Introduction}
	Coronavirus disease 2019 (Covid-19) is an infectious disease caused by the acute respiratory syndrome coronavirus 2 (SARS-CoV-2) \cite{WHOnd2020}. This disease was first reported in 2019 in Wuhan, China, and since then, it has spread globally \cite{hui2020} including in Indonesia. Indonesia confirmed for the first time this Covid-19 case on Monday 2 March 2020 \cite{news2020}.

	General indications of Covid-19 are fever, shortness of breath, cough, muscle aches, diarrhea, and sore throat \cite{patel2020}. Covid-19 is usually spread through the air produced during coughing by one person and can also be spread from touching surfaces contaminated by the virus \cite{patel2020}. Covid-19 can survive on surfaces for up to 72 hours (Health, 2020) dan and the time from exposure to onset of indication is generally two to fourteen days \cite{patel2020}.

	Covid-19 is a global pandemic until 30 April 2021. Based on data published by the World Health Organization (WHO), there are 151517785 confirmed cases from 216 countries \cite{WHO202}. This global pandemic has made many countries, one of which is Indonesia, participate in trying to participate in overcoming it is spread \cite{setiawan2020},\cite{hamid2020}.
	
	Covid-19 is an infectious disease that has the potential to cause public health emergencies. Therefore, the Indonesian Government issued urgency about the formation of regulations related to the prevention of Covid-19 in a Government Regulation and Regulation of the Minister of Health \cite{telaumbanua2020}.

	Based on data published by Indonesian Gugus Tugas Percepatan Penanganan COVID-19 on April 30, 2021, in Indonesia, as many as 1668368 confirmed positive for Covid-19 (COVID, 2020). Because this disease spreads very quickly in a short time, it makes sense to carry out research and analysis on forecasting the number of COVID-19 cases in the future. Then the generalized autoregressive moving average ii modeling results hope to be a recommendation for handling Covid-19 cases based on positive confirmed Covid-19 data in Indonesia.
	
	Forecasting is processing past data to obtain future data estimates \cite{wulandari2017}. There are many types of data available such as interval, nominal, and count data. One of the count data is data from a case. Count data is the data that most often cause data not to spread usually. Therefore, time series forecasting modeling is developed, one of which is arithmetic series data \cite{benjamin2003}.
Modeling is not only closely related to forecasting but also closely related to prediction. The prediction itself is something associated with the outcome of a possibility. Forecasting discusses the analysis of past data to obtain future options, while prediction is an analysis of future cases. The models used in this research are autoregressive moving average and generalized autoregressive moving average.
The autoregressive moving average model consists of two components: Autoregressive and Moving Average. Autoregressive models the autocorrelation of time series variables that depend linearly on the values of the previous variables—Moving Average models the autocorrelation of earlier errors in the time series \cite{hanke2009}. The autoregressive moving average model is a combination of the autoregressive and moving average models so that it assumes that previous data influence the current period data and last error values \cite{kasanah2016}.

	Benjamin et al. developed a generalized autoregressive moving average model for data that follows a non-Gaussian distribution. The generalized autoregressive moving average model is a development of the expansion of generalized linear models (GLM) \cite{hanke2009} and the result of combining autoregressive moving average components with predictor variables to transform the average parameters of the data distribution using the link function \cite{benjamin2003}, \cite{hanke2009}.
In this study, modeling was carried out using the generalized autoregressive moving average. The data used is data on the number of positive cases of Covid-19 in Indonesia from the first confirmed cases of Covid-19 in Indonesia from March 2, 2020, to April 30, 2021.	

%% 	section 2
\section{Literature Review}
\subsection{Autoregressive (AR)}
The autoregressive model $AR\left ( \textit{p} \right )$ is a model which states that data in the current period is influenced by data in the previous period \cite{lilipaly2014}. This model is also the most basic model for a stationary process, or it can also interpreting as a process of regression results by itself. Mathematically it can be written (Asriawan, 2014), \cite{habibi2019}:
\begin{eqnarray}\label{form1}
X_{t}= phi _{1}X_{t-1}+\phi _{2}X_{t-2}+...+\phi _{p}X_{t-p}+\phi _{t}.
\end{eqnarray}
The $X_{t}$ is data in $t$ period with $t=1,2,3,…,n$, $X_{t-i}$ is data in $t-i$ period with $i=1,2,3,…,p$, $\phi _{t}$ is an error in $t$ period, and $\phi _{i}$ is AR coefficient with $i=1,2,3,…,p$.

\subsection{Moving Average (MA)}
Mathematically, the Moving Average $MA\left ( \textit{q} \right )$ model can be written as follows:
\begin{eqnarray}\label{form2}
X_{t}=e _{t}-\theta_{1}e_{t-1}-\theta_{2}e_{t-2}-...-\theta_{q}e_{t-q}.
\end{eqnarray}
The $X_{t}$ is the predicted variable, $\theta_{1},\theta_{2},...,\theta_{q}$ is parameters of Moving Average, and $e _{t},e _{t-1},e _{t-2},...,e _{t-q}$ is error value in $t$ period with $t-1,…,t-q$.

\subsection{Autoregressive Moving Average (ARMA)}
The autoregressive moving average is a model known as the autoregressive integrated moving average model without the ARIMA differentiation process or $ARIMA\left ( p,0,q \right )$. Mathematically $ARMA\left ( p,q \right )$ can be written from \ref{form1} dan \ref{form2}.
\begin{eqnarray}\label{form3}
X_{t}=phi _{1}X_{t-1}+...+\phi _{p}X_{t-p}+e_{t}-\theta _{1}e_{t-1}-...-\theta _{q}e_{t-q}.
\end{eqnarray}
The $X_{t}$ is data in $t$ period with $t=1,2,3,…,n$, $\phi _{1},\phi _{2},...,\phi _{p}$ is parameters of Autoregressive, $\theta_{1},\theta_{2},...,\theta_{q}$ is parameters of Moving Average, and $e _{t},e _{t-1},e _{t-2},...,e _{t-q}$ is error value in $t$ period with $t-1,…,t-q$.
The autoregressive moving average model in time series data applies the Box-Jenkins procedure. The first is model identification. Model identification in autoregressive moving average can see from the autocorrelation function and partial autocorrelation function plots used to determine the p and q orders in the $ARMA\left ( p,q \right )$ model. The second is parameter estimation. Parameter estimation of the model can use the maximum likelihood estimation (MLE) method. The third is forecasting. Forecasting is the final stage of the analysis using time series data. Then, diagnostic check. And the last is the accuracy model.

\subsection{Autocorrelation Function (ACF) and Partial Autocorrelation Function (PACF)}
The autocorrelation function is a linear relationship indicator on observations. For example, the observations $Z_{t}$ and $Z_{t+k}$. While the partial autocorrelation function is an indicator of the magnitude of the relationship between the values of the same variable with other time delays that have an effect that is considered constant.
Autocorrelation function and partial autocorrelation function have two patterns, namely cut off and dies down. When the autocorrelation function and partial autocorrelation function lines are significant in the first lag, but in the subsequent lag, there are not many autocorrelation functions and partial autocorrelation function lines, the pattern is included in the cut off pattern. Meanwhile, if the autocorrelation function and partial autocorrelation function lines decrease gradually or are not cut, the pattern is included in the dies down pattern.
Suppose the autocorrelation function is $\hat\rho{_{k}}$, then:
\begin{eqnarray}\label{form4}
\hat\rho{_{k}}=\frac{\sum_{t=1}^{n-k}\left ( Z_{t}-Z \right )\left ( Z_{t+k}-Z\right )}{\sum_{t=1}^{n-k}\left ( Z_{t}-Z^{} \right )^{2}}.
\end{eqnarray}
And partial autocorrelation function is $\hat\phi{_{kk}}$ with $j=1,2,3,…,k-1$ and $\hat\phi{_{kj}}=\hat\phi{_{k-1,j}}-\hat\phi{_{kk}}\hat\phi{_{k-1,k-j}}$, then:
\begin{eqnarray}\label{form5}
\hat\phi{_{kk}}= \frac{r_{k}-\sum_{j=1}^{k-1}\hat\phi _{{k-1,j}}r_{k-j}}{1-\sum_{j=1}^{k-1}\hat\phi _{{k-1,j}}r_{j}}.
\end{eqnarray}

\subsection{Generalized Autoregressive Moving Average (GARMA)}
The generalized autoregressive moving average model was first introduced by Benjamin \cite{benjamin2003}:
\begin{eqnarray}\label{form6}
g\left ( \mu  \right )=
Z_{t-1}^{T}\beta= X_{t-1}^{T}\beta +\tau _{t}.
\end{eqnarray}
With
\begin{eqnarray}\label{form7}
\tau _{t}= \sum_{j=1}^{p}\phi _{j}A\left ( y_{t-j},x_{t-j},\beta\right )+\sum_{j=1}^{q}\theta _{j}M\left ( y_{t-j},\mu _{t-j}\right ).
\end{eqnarray}
The $\tau _{t}$ is components of AR and MA, $A$ is a function that represents the Autoregressive form, $M$ is a function that represents the moving average form, $\phi^{T}$ is Autoregressive parameter with $\phi^{T}=\left ( \phi _{1},\phi _{2},...,\phi _{p} \right )$, and $\theta^{T}$	is Moving Average parameter $\theta^{T}=\left ( \theta_{1},\theta_{2},...,\theta_{q} \right )$.

\subsection{Maximum Likelihood Estimation (MLE)}
Maximum likelihood estimation is one method that functions as an estimation method that maximizes the likelihood function. Mathematically it can be written \cite{benjamin2003}, (Asriawan, 2014):
\begin{eqnarray}\label{form8}
L\left ( \theta  \right )= \prod_{t=1}^{n}f\left ( x_{i}\mid \theta  \right ).
\end{eqnarray}
Calculating the maximum likelihood estimation can be made easier by using the log-likelihood function, mathematically can be written:
\begin{eqnarray}\label{form9}
l\left ( \theta  \right )= ln L\left ( \theta  \right )=\sum_{i=1}^{n}lnf\left ( x_{i}\mid \theta  \right ).
\end{eqnarray}

\subsection{Validation}

\subsubsection{Mean Absolute Error (MAE) and Root Mean Square Error (RMSE)}
Mean absolute error and root mean square error are validations that are often used to compare and evaluate the accuracy of the results of a prediction from an equation with actual data. Suppose the $(n)$ data with $(e_{i})$ is the error of the model, which is the result of subtraction from the predicted value $(x_{i})$ with the actual value $(y_{i})$ and $i=1,2,3,…,n$, then the mean absolute error value is:
\begin{eqnarray}\label{form10}
MAE=\frac{1}{n}\sum_{i=1}^{n}\left | x_{i}-y_{i} \right |=\frac{1}{n}\sum_{i=1}^{n}\left | e_{i} \right |.
\end{eqnarray}
And the root mean square error value is:
\begin{eqnarray}\label{form11}
RMSE=\sqrt{\frac{1}{n}\sum_{i=1}^{n}\left | e_{i} \right |}.
\end{eqnarray}

\subsubsection{Aikaike's Information Criteria (AIC)}
Aikake's information criterion was first introduced in 1973 by Aikake. Value Aikake's information criterion serves as a conduit of relative size suitability pointer statistical models. In short, the value of Aikake's information criterion is an indicator of the size of the fit of the model with the data being modeled \cite{ray2015}. In addition, the value of Aikake's information criterion also minimizes the loss of information in model selection and provides information to determine the best model. Where the best model obtained is obtained based on the smallest value of Aikake's information criterion \cite{ray2015}.
Value Aikake's information criterion can be obtained by are the parameters of the model and is the maximum value of the  likelihood function that is used to estimate the model and defined as\cite{acquah2010},\cite{anderson1998}, \cite{ray2015}:
\begin{eqnarray}\label{form11}
AIC= 2k-2ln\left ( L \right ).
\end{eqnarray}

%% 	section 3
\section{Result and Discussion}
\begin{figure}[h]
	\centering
	\includegraphics[width=5cm]{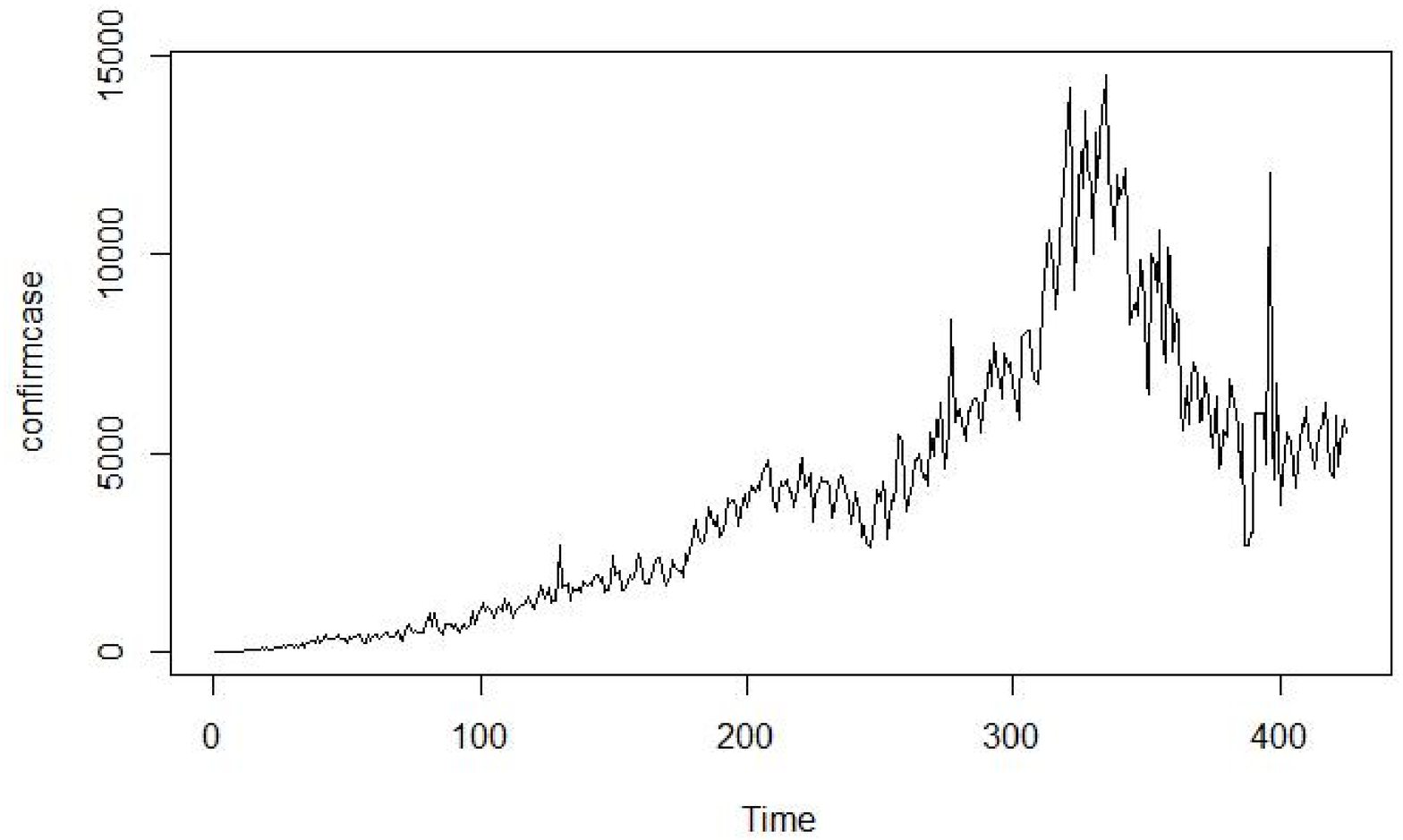}
	\caption{\label{Fig1}Time Series Plot of Data Positive Cases of COVID-19 Per Day in Indonesia from March 2, 2020, to April 30, 2021}
\end{figure}

This study took secondary data from the official website of the task force for the acceleration of handling Covid-19 in Indonesia. The data to be processed is data on positive cases of Covid-19 per day in Indonesia from March 2, 2020, to April 30, 2021, with 425 data. This data has a plot in the time series data presented in Figure \ref{Fig1}.

The first stage in formulating an autoregressive moving average model is identifying data according to the first stage in data analysis. The descriptive statistics from the data are attaching in Table \ref{Tab2}.
\begin{table}[h]%
\centering%
\caption{\label{Tab2}Descriptive Statistics of Data Positive Cases of COVID-19 Per Day in Indonesia from March 2, 2020, to April 30, 2021}
\begin{tabular}{ccc}
\toprule %
\hline
No & Statistic Descriptive of Data & Value    \\ \hline
1  & Min.                          & 0        \\ \hline
2  & $1^{st}$ Quartil              & 1041     \\ \hline
3  & Median                        & 3622     \\ \hline
4  & Mean                          & 3926     \\ \hline
5  & $3^{rd}$ Quartil              & 5803     \\ \hline
6  & Max.                          & 14518    \\ \hline
7  & Sd.                           & 3344.91  \\ \hline
8  & Var.                          & 11188420 \\ \hline 
\bottomrule %
\end{tabular}
\end{table}

The second stage of data analysis is dividing the data into train data and test data. Train data will be processed using the ARMA model with a ratio of 90\% of the data used as train data, and 10\% of the data will be used as test data. Then the divided data, namely the train data, is tested for stationary. In this study, the stationary using Augmented Dickey-Fuller test. The augmented dickey-fuller test has a hypothesis if the $p-value > \alpha$  shows that the data is not stationary or rejects the initial hypothesis, so there is a need to do a differences process. If $p-value <  \alpha$ shows that the data is stationary, there is no need to do a differences process.  In this study, $\alpha$ used is $\alpha = 0.05$ or 5\%. The augmented dickey-fuller test from the data, which has a $p-value$ or $p-value = 0.829> \alpha=0.05$, means the data is not yet stationary or rejects the initial hypothesis, so there is a need to do a differences process.
The augmented dickey-fuller test from the data, after differences process, has a $p-value$ or $p-value=0.01 < \alpha=0.05$, which means the data is stationary or accepts the initial hypothesis, so there is a need to do a differences process.

\begin{figure}[h]
	\centering
	\includegraphics[width=5cm]{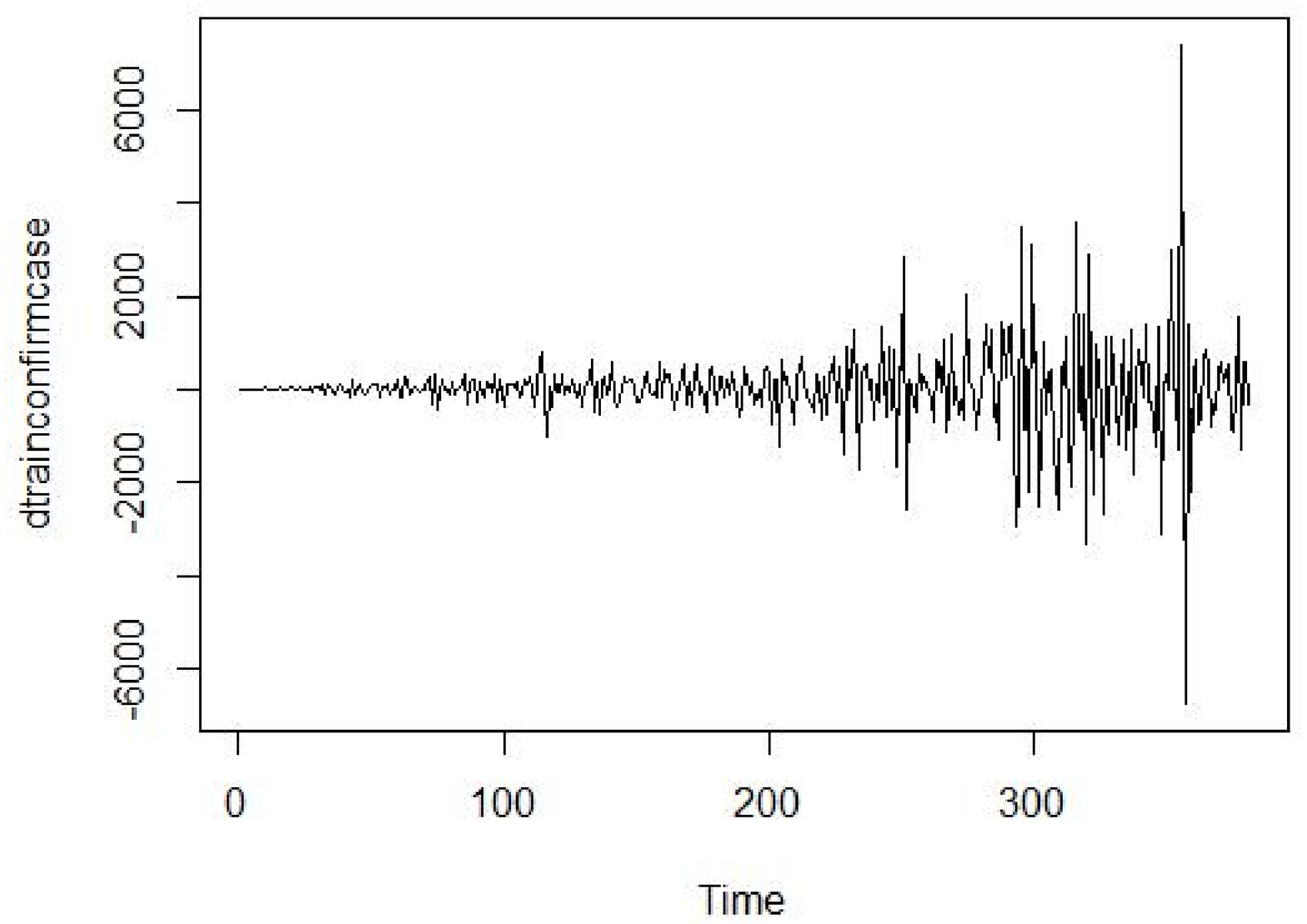}
	\caption{\label{Fig2}Time Series Plot of Stationary Data Positive Cases of COVID-19 Per Day in Indonesia}
\end{figure}

\begin{figure}[h]
	\centering
	\includegraphics[width=5cm]{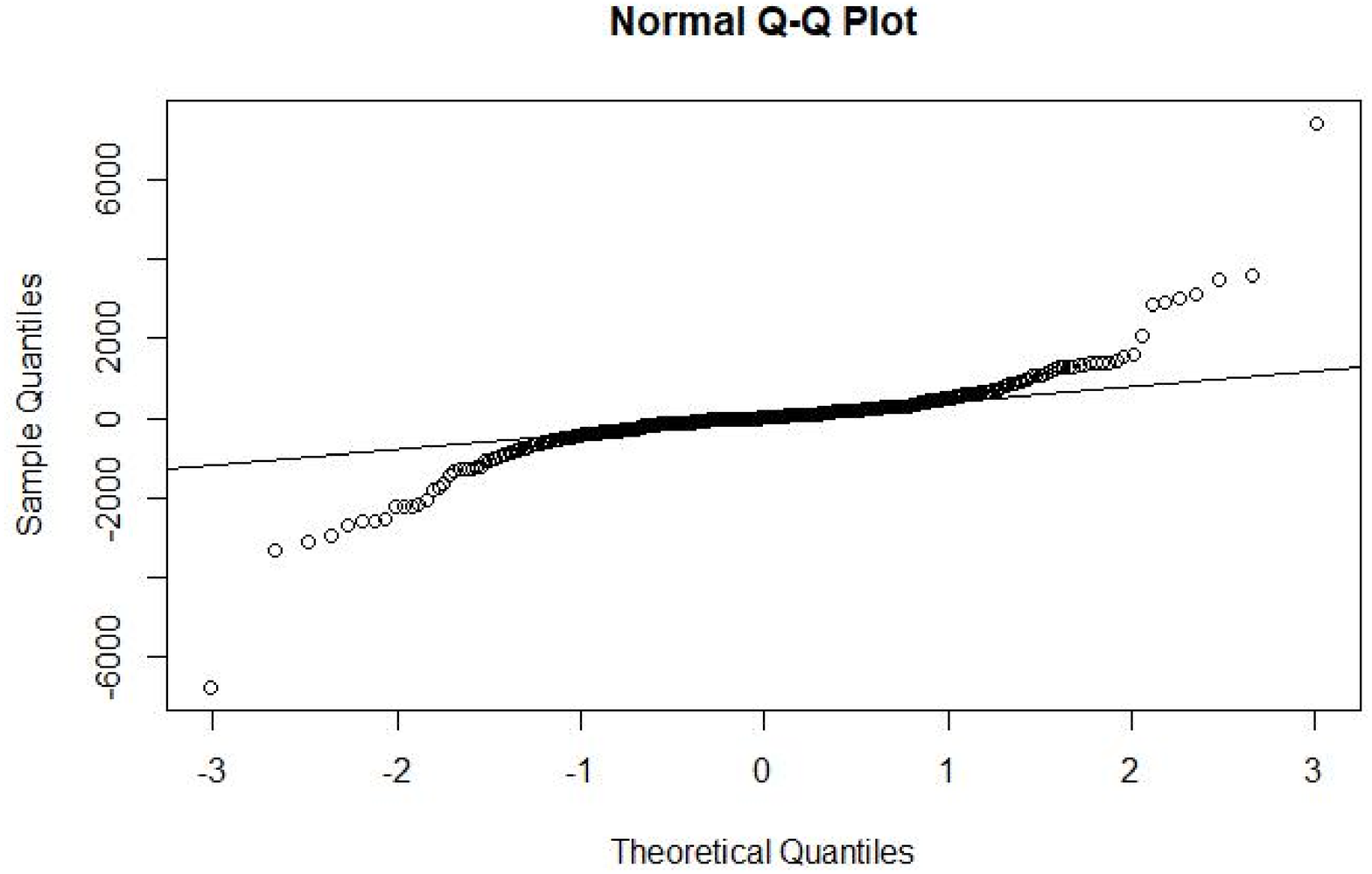}
	\caption{\label{Fig3}Q-Q Plot of Stationary Data Positive Cases of COVID-19 Per Day in Indonesia}
\end{figure}

Time series plot of stationary data is attaching on Figure \ref{Fig2} and The stationary time series data can be seen from the qqnorm plot, which is linked in Figure \ref{Fig3}.

\begin{figure}[h]
	\centering
	\includegraphics[width=5cm]{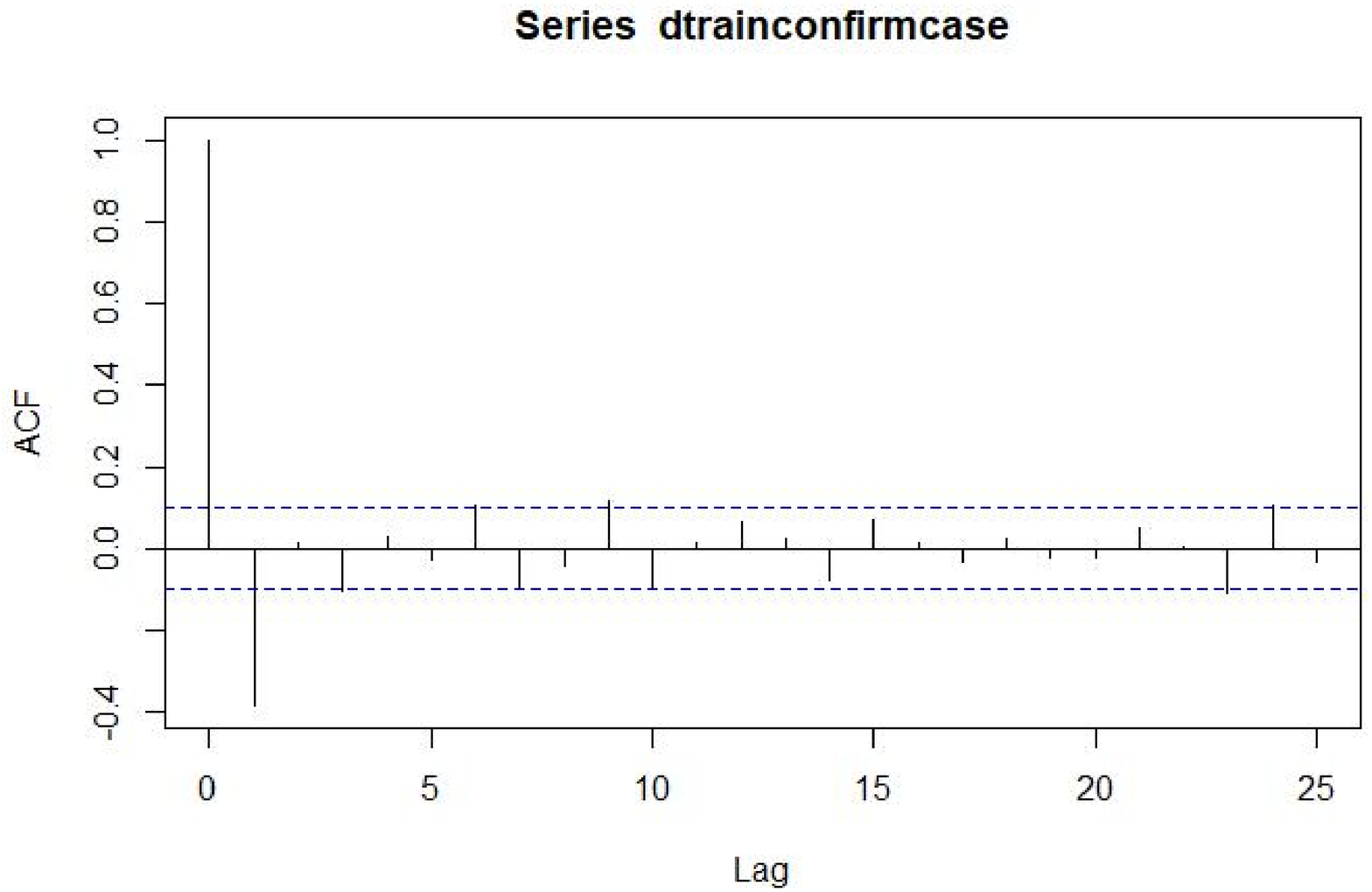}
	\caption{\label{Fig4}Autocorrelation function Plot of Stationary Data Positive Cases of COVID-19 Per Day in Indonesia}
\end{figure}

\begin{figure}[h]
	\centering
	\includegraphics[width=5cm]{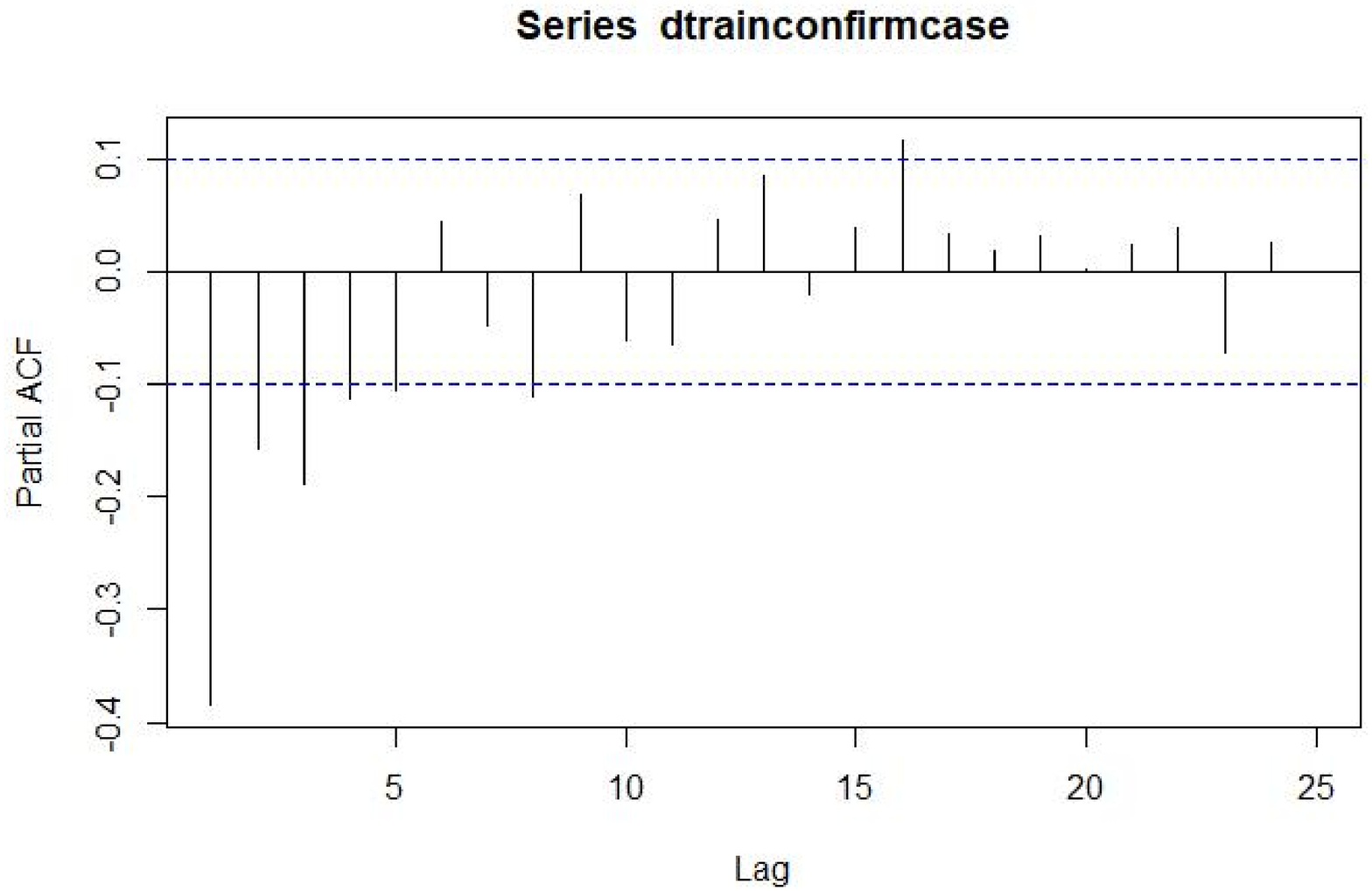}
	\caption{\label{Fig5}Partial autocorrelation function Plot of Stationary Data Positive Cases of COVID-19 Per Day in Indonesia}
\end{figure}

The third step in data analysis is forecasting the ARMA model. The first step in forecasting the ARMA model is identifying the model. The aim of identifying the model is to determine the order in the ARMA model. The order is p and q, obtained from the autocorrelation function and partial autocorrelation function plots. The autocorrelation function and partial autocorrelation function plots are attaching in Figure \ref{Fig4} and \ref{Fig5}.

\begin{figure}[h]
	\centering
	\includegraphics[width=5cm]{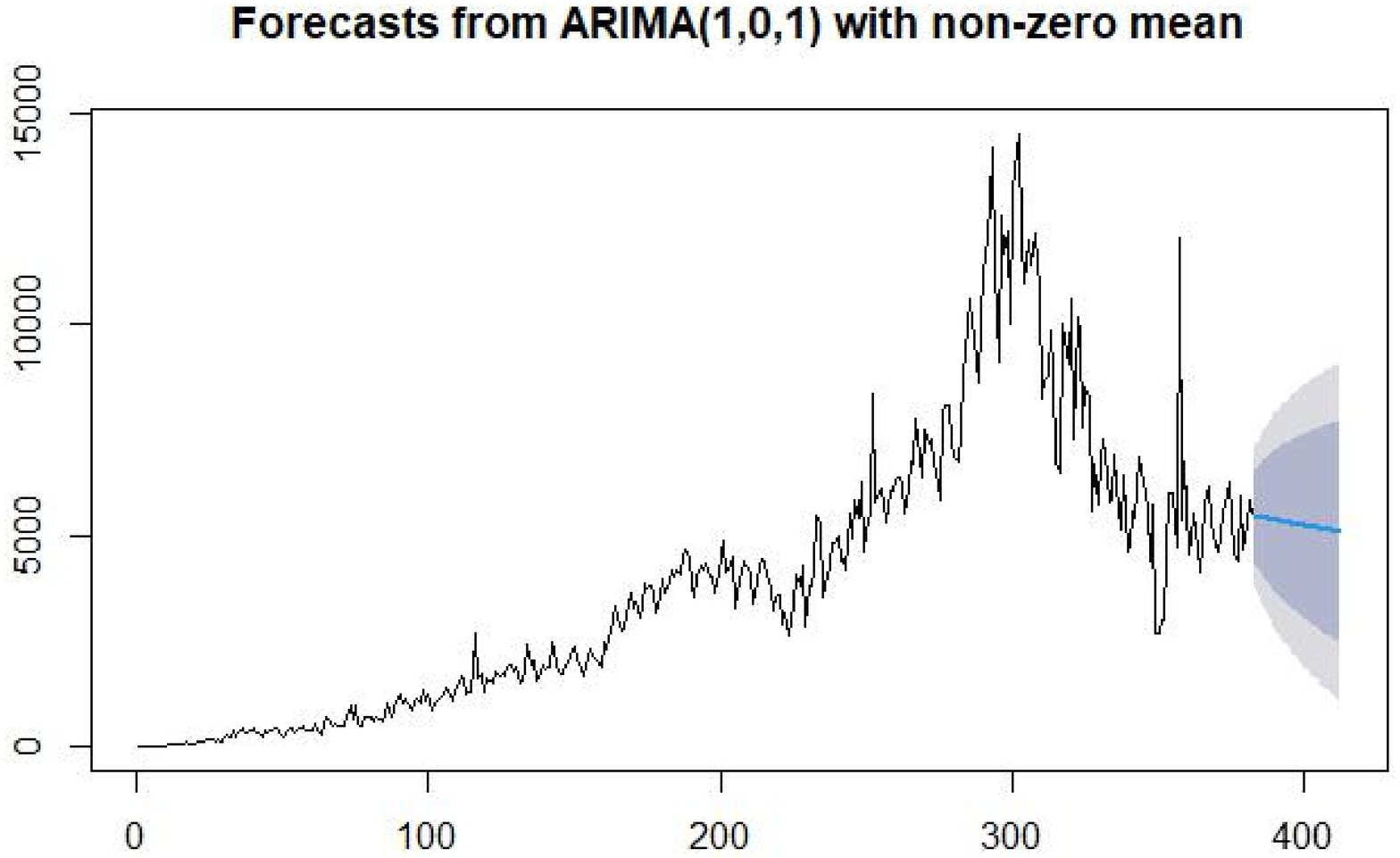}
	\caption{\label{Fig6}Forecasting Plot of ARMA(1,1) Model}
\end{figure}

\begin{figure}[h]
	\centering
	\includegraphics[width=5cm]{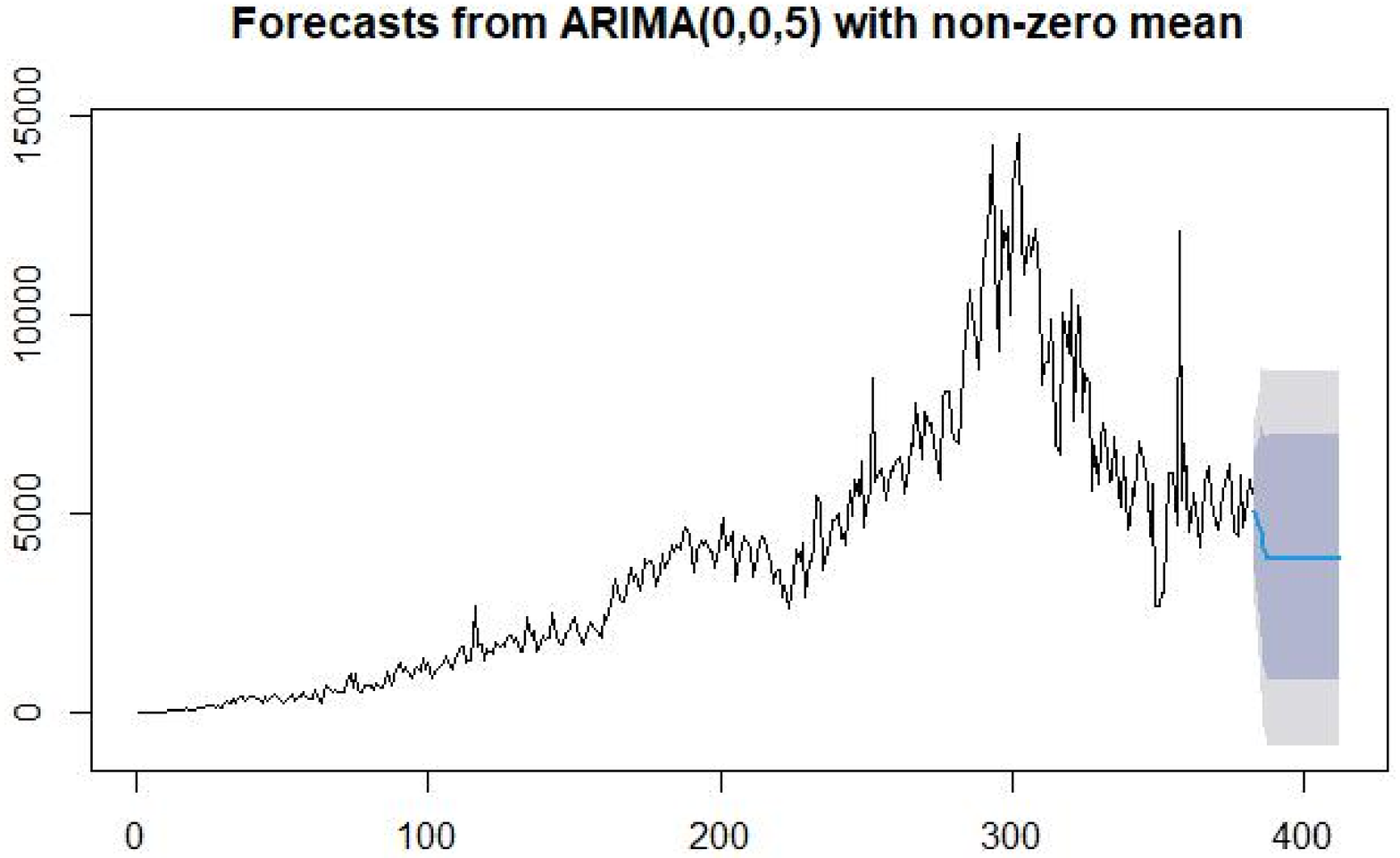}
	\caption{\label{Fig7}Forecasting Plot of MA(5) Model}
\end{figure}

Based on the autocorrelation function and partial autocorrelation function from Figure \ref{Fig4} and \ref{Fig5} a model formed from data on positive cases of Covid-19 per day in Indonesia from March 2, 2020, to April 30, 2021, is an ARMA(1,1) and MA(5) and the model also be obtained by ARIMA(1,0,1) and ARIMA(0,0,5). From the ARMA(1,1) and MA(5) models, the forecasting results were obtained having an RMSE value of $850,654$, MAE of $497,2133$, and an AIC value of $6249$ for the ARMA(1,1) model and the MA(5) model having RMSE value is $1193,662$, MAE is $820,0247$, and AIC is $6512.66$ forecasting results with the ARMA(1,1), and MA(5) has a plot as in Figure \ref{Fig6} and Figure \ref{Fig7}.

The modeling results are ARMA(1,1) and MA(5) copy can be used for forecasting can be used for prediction. Figures 8 and 9 show the prediction results of the ARMA(1,1) and MA(5) models against the original data. This prediction result has a red plot and a black color for the original data.
\begin{figure}[h]
	\centering
	\includegraphics[width=5cm]{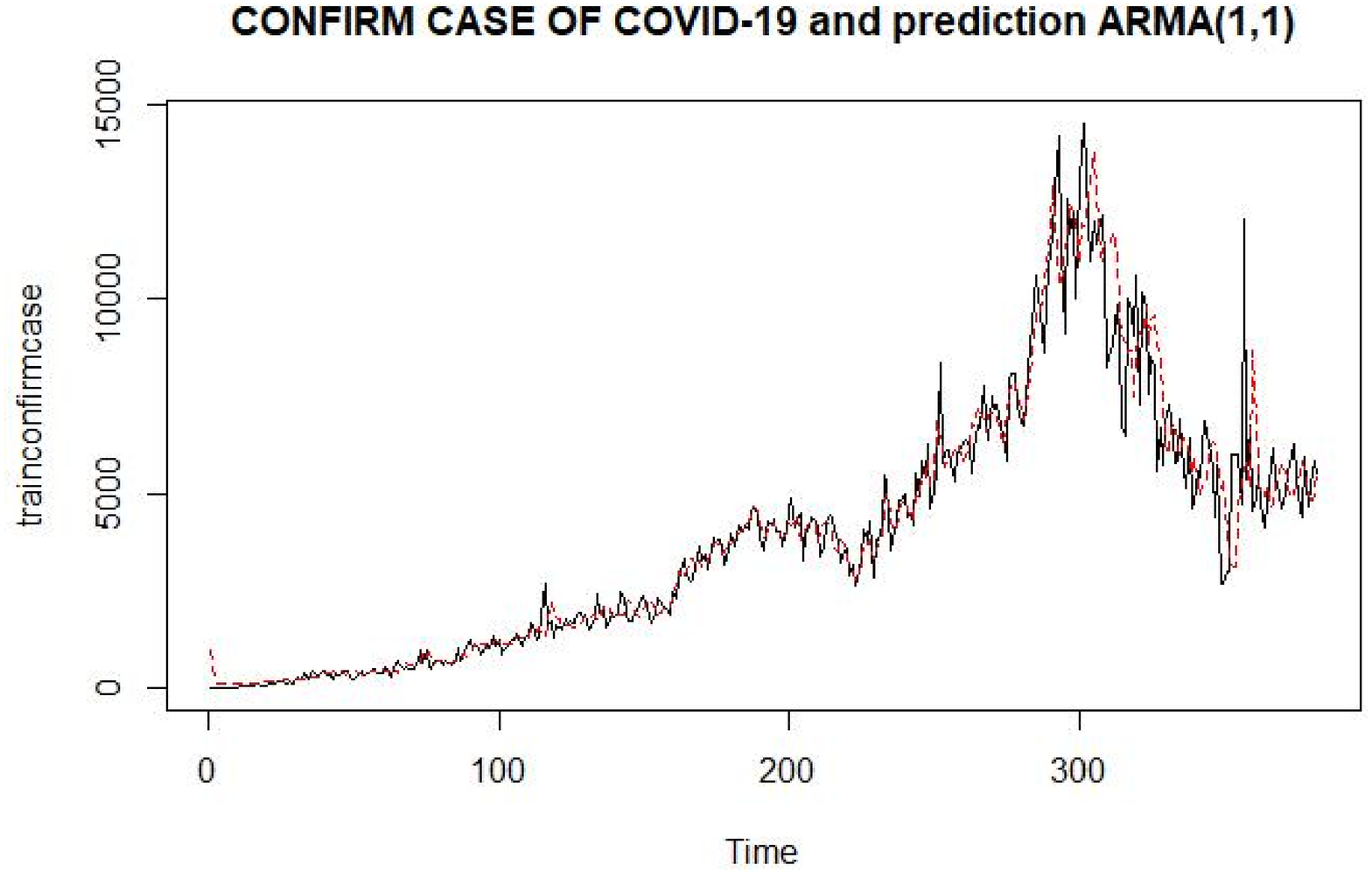}
	\caption{\label{Fig8}Prediction Plot of ARMA(1,1) Model}
\end{figure}
\begin{figure}[h]
	\centering
	\includegraphics[width=5cm]{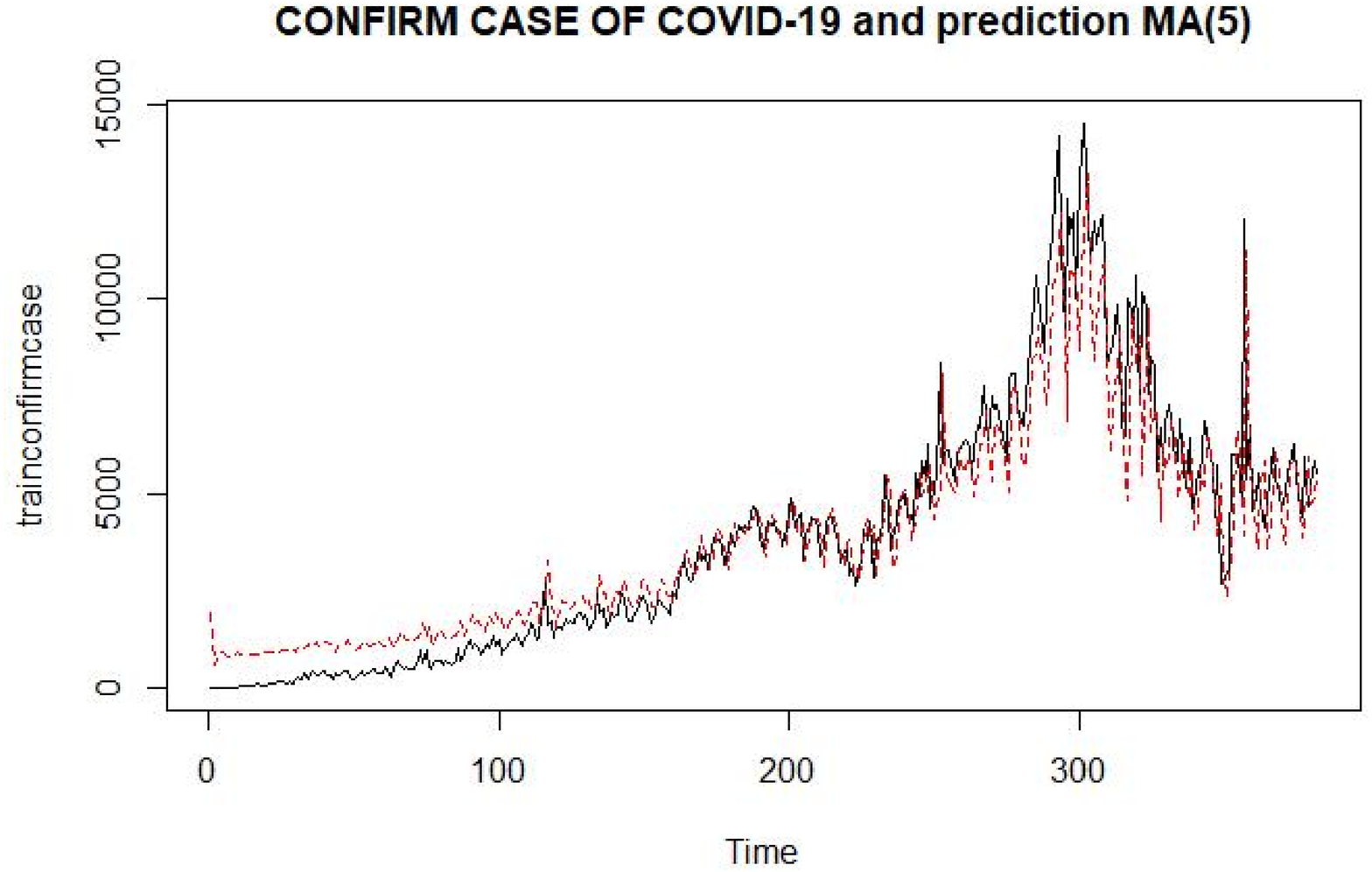}
	\caption{\label{Fig9}Prediction Plot of MA(5) Model}
\end{figure}

Based on the error values, namely the Root Mean Square Error (RMSE), Mean Absolute Error (MAE), and AIC values and the results of forecasting and predictions, it is obtained that the best model for data on positive cases of COVID-19 in Indonesia is ARMA(1,1).
The residual ARMA(1,1) model has checked by diagnostic check uses the Ljung-Box method with the hypothesis that if $p-value>\alpha $ indicates that the residual is independently distributed and if the $p-value< \alpha $ means that the residual of the model not independently distributed (serial correlation). The result of the Ljung Box of ARMA(1,1) model has $0.04591$,  $\alpha$ used is equal to  $\alpha =0.05$ or 5\%, which means that the null hypothesis is rejected. We can conclude that the residuals are independently distributed, or it shows that with ARMA(1,1) model, the assumption of making a mistake is 5\%. The residual of the ARMA(1,1 ) model is independent at the 95\% level. ARMA(1,1) model is a good fit model and has a residual result:
\begin{figure}[h]
	\centering
	\includegraphics[width=5cm]{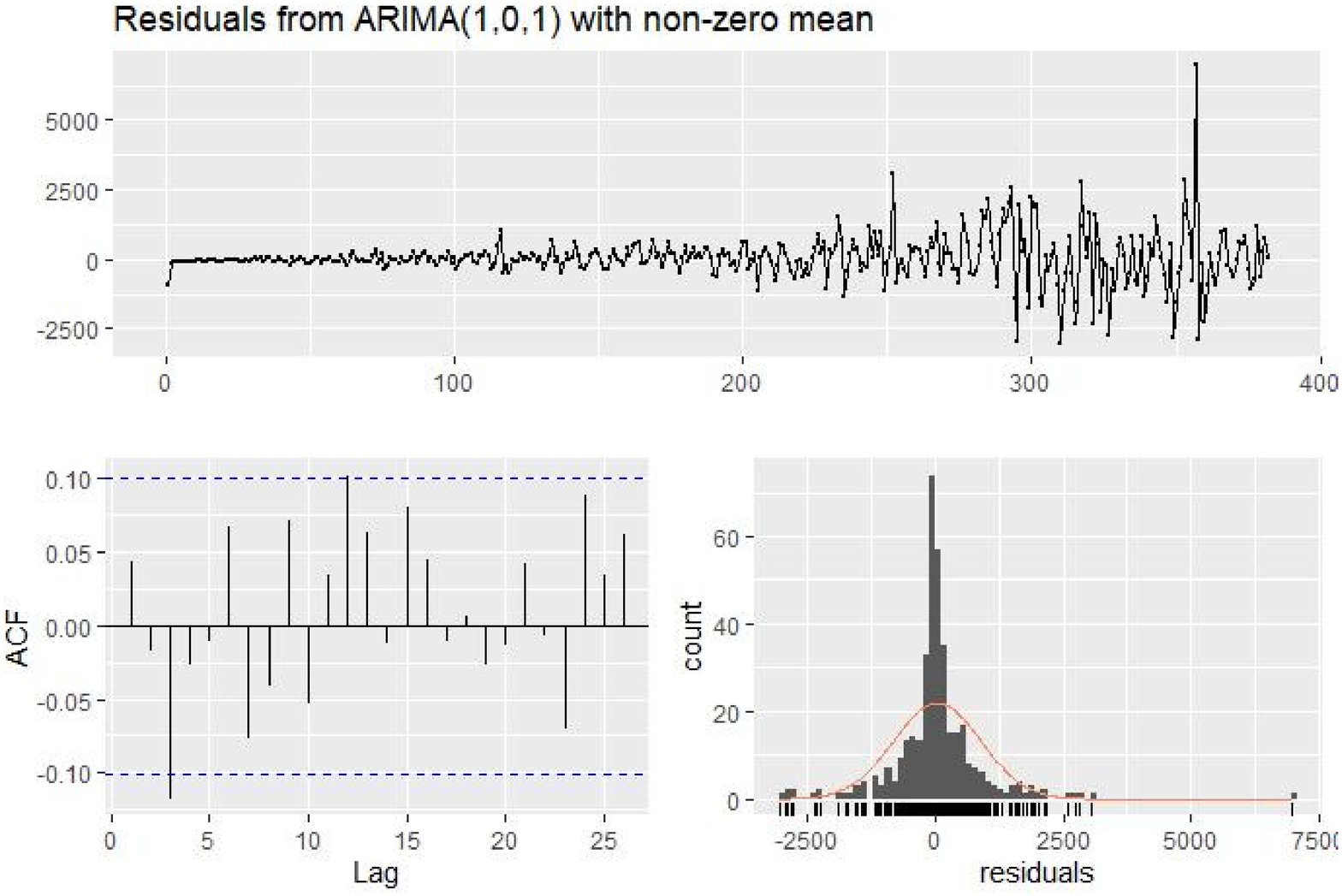}
	\caption{\label{Fig10}The Residual Plot of ARMA(1,1) Model}
\end{figure}

The best previous modeling result based on the smallest error value obtained is ARMA(1,1). This ARMA(1,1) model will then be used for calculations on GARMA with the GARMA(1,1) model calculated. The results of GARMA(1,1) modeling are obtained:
\begin{table}[h!]%
\centering%
\caption{\label{Tab3}The results of modeling with the GARMA(1,1) model}
\begin{tabular}{cccc}
\toprule %
\hline
No & \multicolumn{2}{c}{estimate} & Std Error \\ \hline
1  & intercept     & -4.64217     & 0.031575  \\ \hline
2  & lag           & 1.719924     & 0.001682  \\ \hline 
\bottomrule %
\end{tabular}
\end{table}

The modeling results, namely through the ARMA model in the form of forecasting and predictions, then the results of the GARMA ii modeling have the hope of being a recommendation for handling Covid-19 cases based on positive confirmed Covid-19 data in Indonesia.
This research focuses on data confirmed to be positive for Covid-19 in Indonesia because from the start of the pandemic until now, positive cases of Covid-19 in Indonesia are still the main focus, and based on data, positive cases of Covid-19 in Indonesia are still increasing.
The modeling in this study includes time series count data that uses two models, namely ARMA and GARMA. Research on time series count data not only uses the ARMA and GARMA models, but some of its developments are also available through ARMA-GARCH, Bayesian GARMA, and others.

%% 	section 4
\section{Conclusions}
Research that focuses on data on positive cases of Covid-19 in Indonesia from March 2, 2020, to April 30, 2021, which are time-series data, was forecasted and predicted to obtain two models. The models used are ARMA and GARMA with the hope of being a recommendation for handling Covid-19 cases based on positive confirmed Covid-19 data in Indonesia. The modeling results obtained are the ARMA(1.1) model with an RMSE value of 850,654, MAE of $497,2133$, AIC value of $6249$ and MA(5) with a value of RMSE of $1193,662$, MAE of $820,0247$, and AIC of $6512.66$. Of the two models, the best model obtained is ARMA(1,1), with a smaller error value than the MA(5) model. The best model obtained by ARMA(1,1) is continued to modeling with GARMA(1,1). The GARMA(1.1) model produces parameter estimates of $-4.64217$ and $1.719924$ with errors of $0.031575$ and $0.001682$, respectively. The modeling results in this study can also be continued through the ARMA-GARCH model and the Bayesian GARMA model.

\section*{Acknowledgements}
We would like to thank the referees for
his comments and suggestions on the manuscript.

%%----------		References		----------%%

\end{document}